\documentclass[12pt]{article}
\usepackage{graphicx}
\begin{document}
A Market of Inhomogeneous Threshold Cellular Automata
\bigskip

Dietrich Stauffer* and G\'erard Weisbuch**

\noindent
*Department of Physics, Indian Institute of Technology, Kanpur 208016, India;
visiting from
Institute for Theoretical Physics, Cologne University, D-50923 K\"oln, Euroland

\noindent
**Laboratoire de Physique Statistique, ENS, 24 rue Lhomond, F-75231 Paris Cedex 5,
Euroland

\bigskip

{\small This article summarizes some physics aspects of the market model of Weisbuch and
Stauffer, Physica A (2003): How do demand and quality expectation adjust to 
each other in a buyer dominated market like cinema visits? And how can business
cycles be modelled? }

\bigskip
``Social percolation'' \cite{socper} applied percolation theory to a marketing
problem: How to quality expectations of potential buyers and the quality of
a product adjust to each other ? The model lead to self-organized criticality
at the percolation threshold. Thus fluctuations were large, some products were
a big hit, most were flops and lots of products were somewhere in the middle.
Fig.1a \cite{proykova} shows one of the simulations, and Fig.1b a corresponding
result from a real market (cinema films; see also \cite{vany}). There is qualitative
agreement; note that the simulation rank plot extends over four decades, the
one with real data only over two decades, in vertical direction.

For this ``threshold'' conference, instead of this percolation model we apply 
threshold cellular automata to this marketing problem: How many potential buyers
actually buy ? Different people have different opinions, and thus our $N$
automata have different thresholds $\theta_i, \; i=1,2, \dots N$. This is what
is meant with ``inhomogeneous''. Each site $i$ on a large square lattice 
carries besides the threshold $\theta_i$ a spin $S_i$ which is +1 for buying
and --1 for not buying. The spin orientation at the next time step is fully
determined by the sum $Q=\sum_kS_k$ of the four neighbour spins $k$ of $i$:
If this ``quality'' $Q$ is larger than the threshold $\theta_i$, site $i$ 
buys; otherwise it does not buy. The magnetization $M = \sum_i S_i/N$ is the
normalized difference between the numbers of buyers and non-buyers. We start
with all spins up and then use random sequential updating of the spins, with
one time step meaning that on average every site was updated once. 

As long as the thresholds do not vary with time, the above problem is similar
to a zero-temperature simulation of a random-field Ising model. However, after
going to a movie you may not go immediately to another one except when it is
very good; in contrast, when you have not seen a good movie since a long time
you may visit even a mediocre one. Thus $Q$ is moved up if $S_i$ now is +1,
and moves down if $S_i = -1$. Since the quality $Q$ as reported from the 
neighbours can take only the values $-4, -2, 0, 2, 4$, a threshold 2.1 is 
equivalent to a threshold 3 or 3.9. Thus for simplicity we take our thresholds
as odd integers $-5, -3, -1, 1, 2, 3$, initially distributed randomly, and
change them at each time step by $2 S_i = \pm 2$. When the threshold has reached
+5 the next updating of this spin will change $S_i$ to --1 and $\theta_i$ to 3,
since $Q$ cannot reach 5. (If one allows the thresholds the change by only a
small amount, the dynamics gets slow and is described by a scaling law
\cite{weisbuch}; we ignore this complication here.)

This model thus describes a market dominated by demand of buyers and assuming
sufficient supply of sellers. The judgment of the buyers is determined by what
their neighbours do; if a sufficient number of neighbours buy, then also the
central person buys. Which number of buying neighbours is sufficient in this 
sense depends on the time-dependent threshold $\theta_i(t)$. This threshold
increases by 2 (i.e. by one more neighbour) whenever the person buys, and 
decreases by the same amount when the person does not buy. 

\bigskip
{\bf Results:}

Fig.2 shows that in a stationary equilibrium the buyers are not randomly 
distributed but form small clusters. The cluster size is much smaller than 
the lattice size (usually one million sites); in contrast to social 
percolation \cite{socper}, we do not observe self-organized criticality. 
The fluctuations in the stationary magnetization decrease as $1/\sqrt N$ as
usual: Fig.3. The magnetization ($M = 1$ initially) decays to zero with
damped oscillations, in one (Fig.4), two (Fig.5), three (Fig.6) and four 
(Fig.7) dimensionsions (hypercubic lattices). (Also the triangular lattice
behaves similarly; not shown). 

When you have drunken one bottle of wine, you should not immediately buy
and drink the next one. Thus a site which just bought should wait $\tau$
time steps before considering to buy again. During this waiting time its
spin stays at its value and counts as buyer. Analogously, after a decision
not to buy, the spin stays at $-1$ for $\tau$ iterations and counts as not
buying, to preserve up-down symmetry. The previous description thus is the
special case $\tau = 0$. Then the oscillations are damped much less, as can be 
seen from Figs. 4-6. The amplitude of the oscillations decays exponentially
with time at intermediate time scales (not shown).

Durable goods like houses are often bought only once in life. Setting
in some sense $\tau = \infty$ but sacrificing the up-down symmetry kept until 
now, we assume that once a site has bought in never buys again. We still start
with all $S_i = 1$ and $\theta_i$ random. Fig.8 then shows the numbers of 
actually still buying people; this number initially may vary non-monotonically
with time but later decays exponentially with time until the whole market 
is saturated forever.

When we go back to the above reversible model with $\tau = 0$ and start with 
randomly oriented spins instead of all spins parallel, then the various clusters
oscillate with different phases, showing oscillations for the overall 
magnetization $M(t)$ only for small lattices. In that case it is better to
observe the autocorrelation function $<M(T+t)M(T)>$ versus $t$ for large times
$T$ such that equilibrium is established; this function then looks similar
to the $M(t)$ of Fig.3. More information, including Fourier transforms, on 
this way of initialization are given in \cite{weisbuch}.

At the conference, many suggestions were made by the audiance, like putting this
model on a Barab\'asi-Albert or small-world network, allowing for quenched or 
annealed solution, tune the model to the percolation threshold. J. Kert\'esz
suggested to put in a fraction $p$ of antiferromagnetic bonds, i.e. mistrust
of neighbours: If this neighbour likes the movie it must be bad. We found that
for $\tau=0$ the oscillations vanish if at least half of the bonds are 
antiferromagnetic, while for $\tau \ge 2$ one overshooting takes place even 
if all bonds are antiferromagnetic. 

DS thanks a DFG/BMZ grant for Indian-German collaboration for supporting this
visit.

%\newpage

\begin{figure}[hbt]
\begin{center}
\includegraphics[angle=-90,scale=0.39]{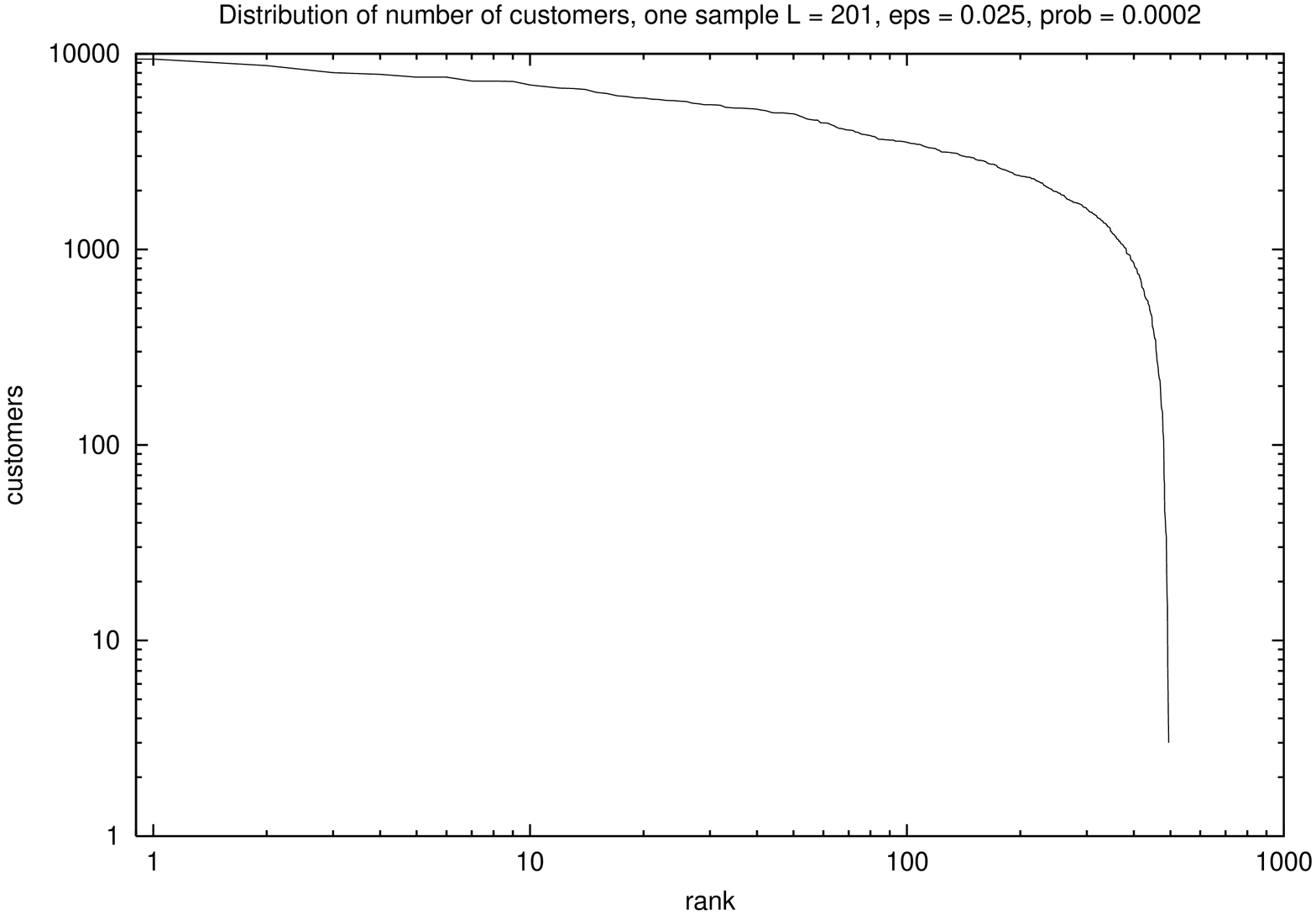}
\includegraphics[angle=-90,scale=0.39]{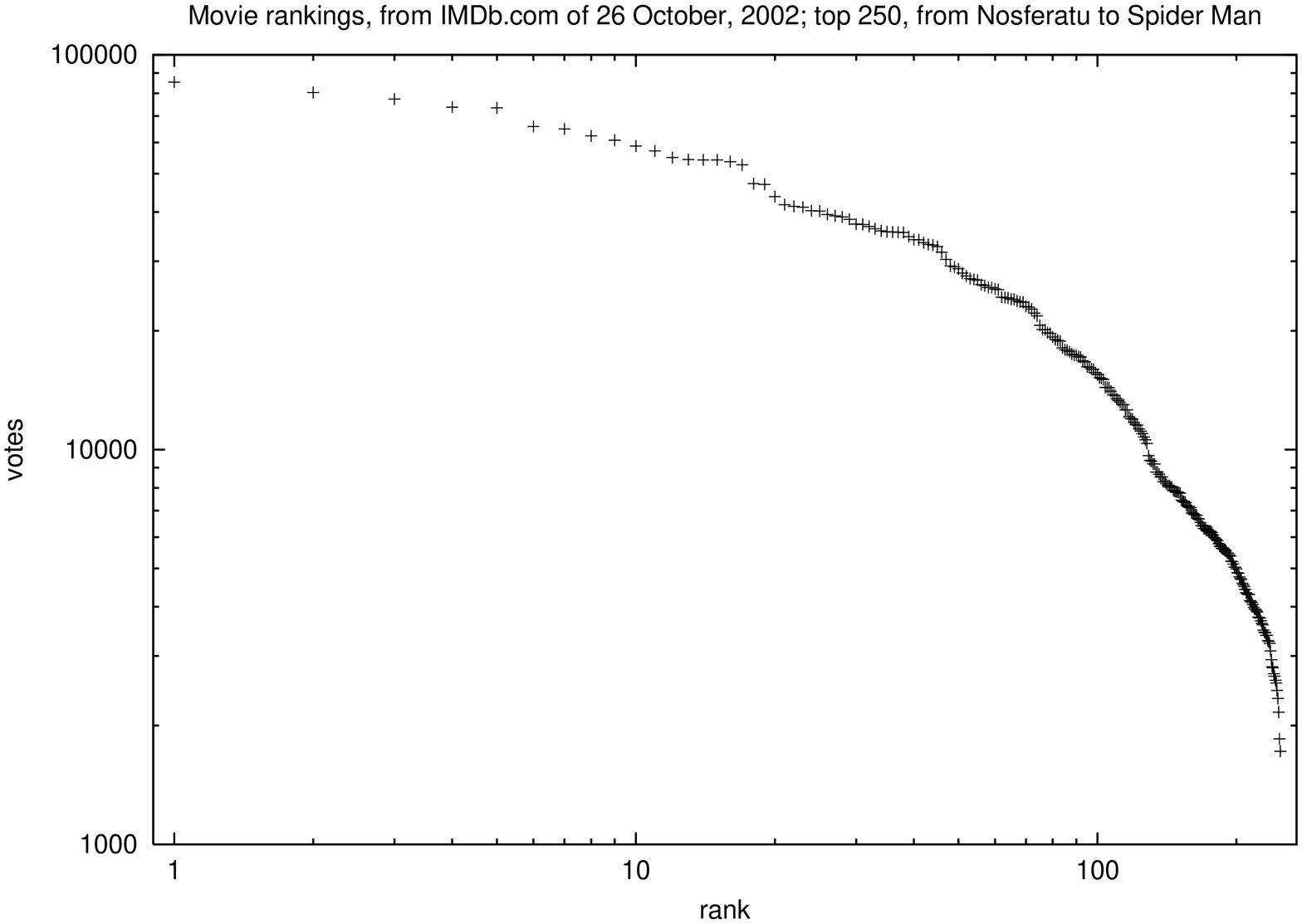}
\end{center}
\caption{
a) Social percolation simulation, with small advertising \cite{proykova}.
We show a rank plot, i.e. the most successful ``movie'' is placed on 
rank 1, the second-most successful one on rank 2, etc.
b) Rank plot of real movies, similar to \cite{vany}, from IMDb.com.
}
\end{figure}

\begin{figure}[hbt]
\begin{center}
\includegraphics[angle=-90,scale=0.6]{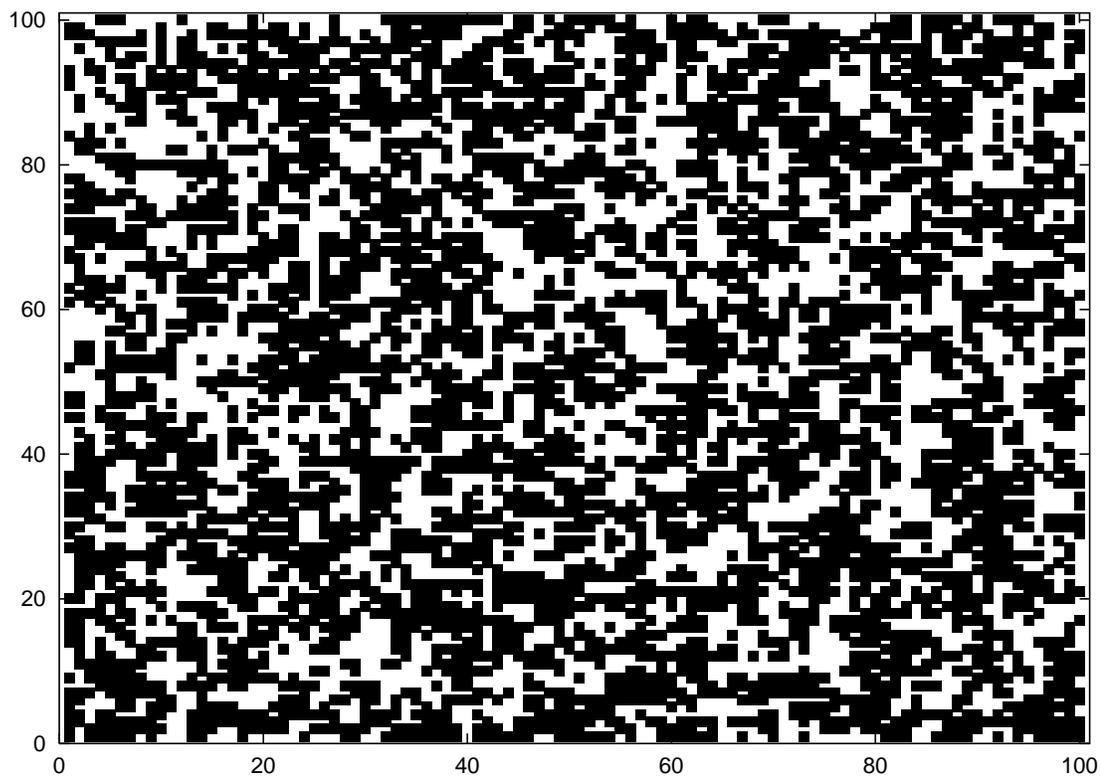}
\end{center}
\caption{
Example of $100 \times 100$ square lattice, $\tau = 0$ after 300 
iterations. Only the up spins are shown.
}
\end{figure}

\newpage

\begin{figure}[hbt]
\begin{center}
\includegraphics[angle=-90,scale=0.39]{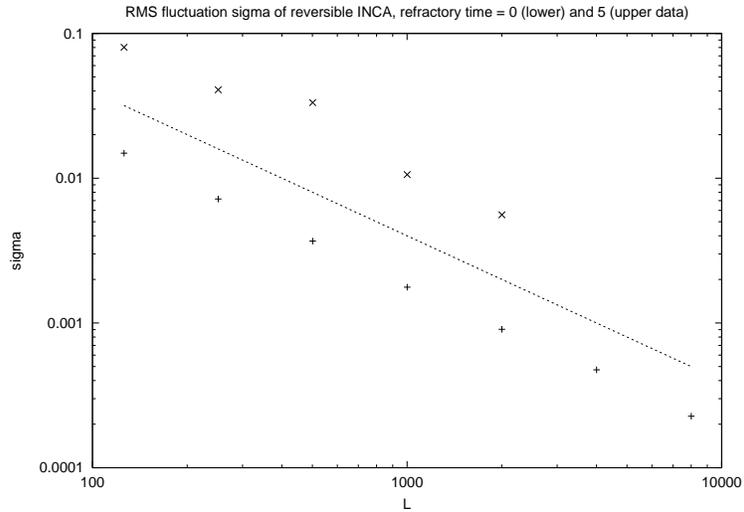}
\end{center}
\caption{
Fluctuations in $M$ versus linear dimension $L$ of square lattice, for $\tau$ 
= 0 (+) and = 5 (x). The straight line in this log-log plot has the slope
--1. 
}
\end{figure}

\begin{figure}[hbt]
\begin{center}
\includegraphics[angle=-90,scale=0.39]{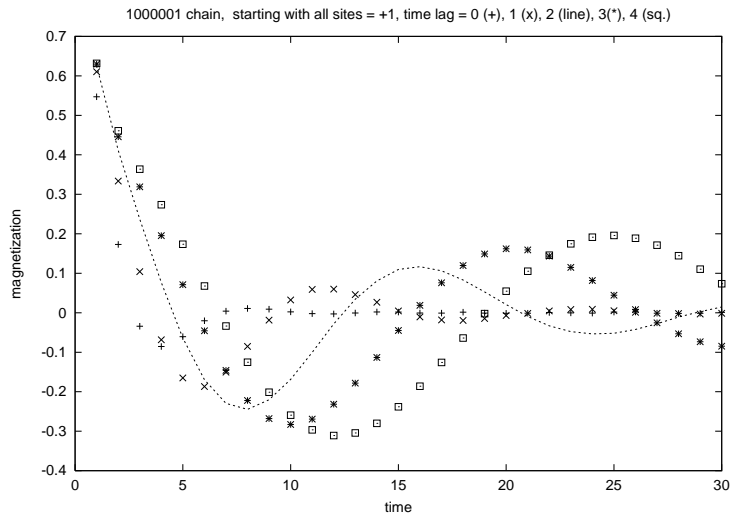}
\end{center}
\caption{
Magnetisation versus time for one dimension, $\tau$ = 0 to 4 as shown in 
headline. 
}
\end{figure}

\begin{figure}[hbt]
\begin{center}
\includegraphics[angle=-90,scale=0.39]{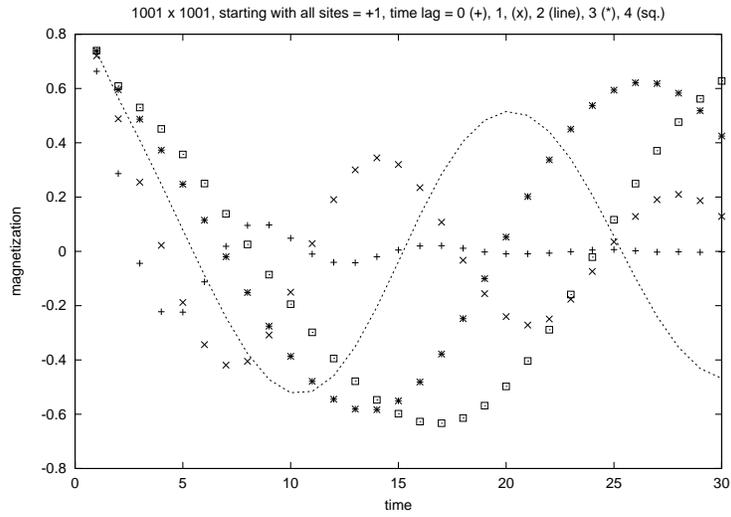}
\end{center}
\caption{
As Fig. 4 but for two dimensions.
}
\end{figure}

\begin{figure}[hbt]
\begin{center}
\includegraphics[angle=-90,scale=0.39]{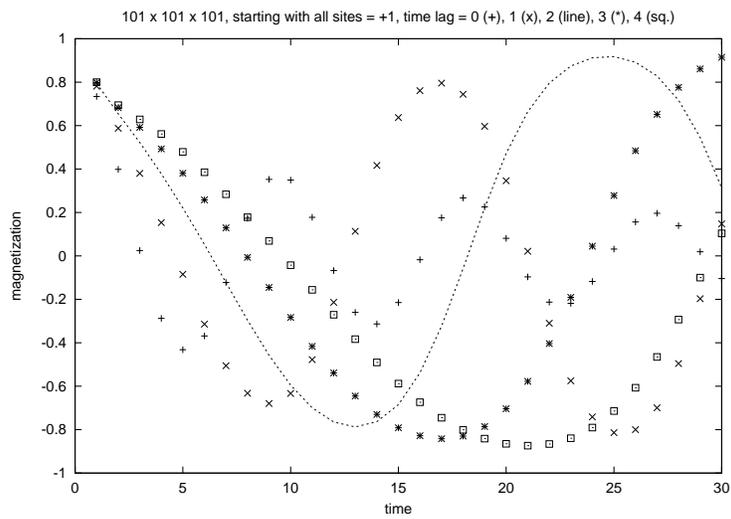}
\end{center}
\caption{
As Fig. 4 but for three dimensions.
}
\end{figure}

\begin{figure}[hbt]
\begin{center}
\includegraphics[angle=-90,scale=0.39]{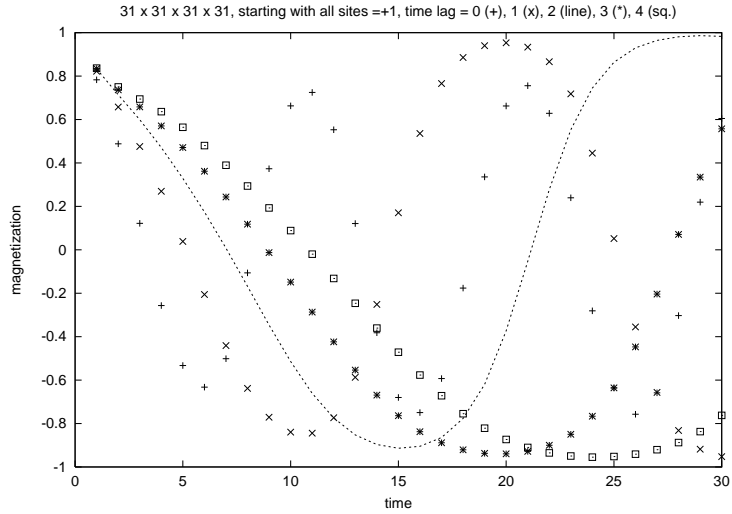}
\end{center}
\caption{
As Fig. 4 but for four dimensions.
}
\end{figure}

\begin{figure}[hbt]
\begin{center}
\includegraphics[angle=-90,scale=0.39]{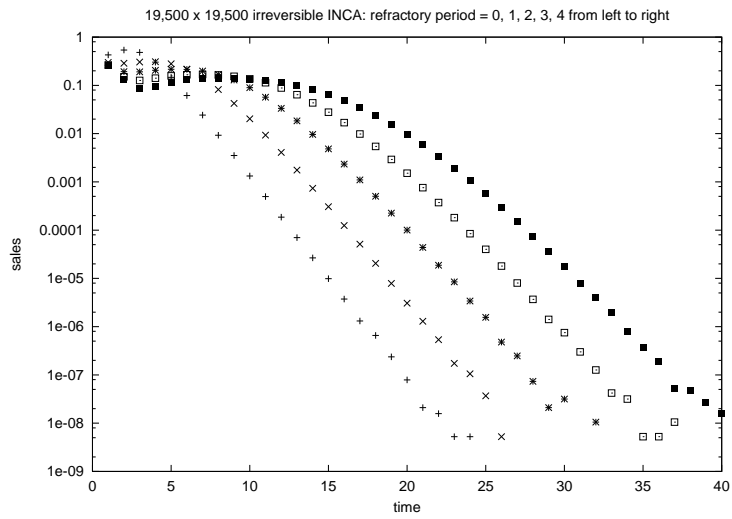}
\end{center}
\caption{
Fraction of sales for 380 million irreversible automata who buy only once.
}
\end{figure}

\end{document}